\documentclass[conference]{IEEEtran}
\usepackage{graphicx}
\usepackage{url}
\usepackage{subfigure}
\usepackage{color}
\usepackage{booktabs}
\usepackage{fontawesome}
\usepackage{amsmath}
\usepackage{multicol} 
\usepackage{multirow} 
\usepackage[ruled]{algorithm2e}

\usepackage{listings}
\usepackage{xcolor}
\definecolor{codegreen}{rgb}{0,0.6,0}
\definecolor{codegray}{rgb}{0.5,0.5,0.5}
\definecolor{codepurple}{rgb}{0.58,0,0.82}
\definecolor{backcolour}{rgb}{0.95,0.95,0.92}

\lstdefinestyle{mystyle}{
    backgroundcolor=\color{backcolour},   
    commentstyle=\color{codegreen},
    keywordstyle=\color{magenta},
    numberstyle=\tiny\color{codegray},
    stringstyle=\color{codepurple},
    basicstyle=\ttfamily\footnotesize,
    breakatwhitespace=false,         
    breaklines=true,                 
    captionpos=b,                    
    keepspaces=true,                 
    showspaces=false,                
    showstringspaces=false,
    showtabs=false,                  
    tabsize=2 
}

\newcommand\fig{Fig.}
\newcommand\tab{Tab.}

\newcommand\snkv{\textit{SNK-Visualizer}~}
\newcommand\snks{\textit{SNK-Scenario}~}
\newcommand\snkser{\textit{SNK-Server}~}
\newcommand\snka{\textit{SNK-Analyzer}~}
\newcommand\sn{SN}
\newcommand\sns{SNs}

\begin{document}

\title{Space Networking Kit: A Novel Simulation Platform for Emerging LEO Mega-constellations}

\author{
	\IEEEauthorblockN{
		Xiangtong Wang\IEEEauthorrefmark{1},
		Xiaodong Han\IEEEauthorrefmark{2},
		Menglong Yang\IEEEauthorrefmark{1}\IEEEauthorrefmark{3},
		Songchen Han\IEEEauthorrefmark{1},
		Wei Li\IEEEauthorrefmark{1}\IEEEauthorrefmark{3}
		}
		
	\IEEEauthorblockA{
		\IEEEauthorrefmark{1}School of Aeronautics and Astronautics, Sichuan University, Chengdu, China
	}
	\IEEEauthorblockA{
		\IEEEauthorrefmark{2}China Academy of Space Technology, Beijing, China
	}

	\IEEEauthorblockA{
		\IEEEauthorrefmark{3}Robotic Satellite Key Laboratory of Sichuan Province, Chengdu, China
	}

	\IEEEauthorblockA{
		Email: \{li.wei,hansongchen\}@scu.edu.cn
	}
}

\maketitle

\begin{abstract}

    This paper presents SNK, a novel simulation platform designed to evaluate the network performance of constellation systems for global Internet services.
    SNK offers real-time communication visualization and supports the simulation of routing between edge node of network. 
    The platform enables the evaluation of routing and network performance metrics such as latency, stretch, network capacity, and throughput under different network structures and density.
    The effectiveness of SNK is demonstrated through various simulation cases, including the routing between fixed edge stations or mobile edge stations and analysis of space network structures.

    \end{abstract}

\IEEEpeerreviewmaketitle

\section{Introduction}

The development of satellite networks (SN) within the "NewSpace" paradigm has gained substantial traction in recent years and holds significant promise in delivering global internet connectivity characterized by low latency and high throughput. Commercial enterprises, such as SpaceX\cite{starlink}, Amazon\cite{kuiper}, Oneweb\cite{oneweb}, etc., have joined this competitive “NewSpace” race. 
While above exciting prospects depict a promising picture of these future integrated satellite-terrestrial networks, the community still has very limited understanding of the structural characteristic and the available network performance of modern mega-constellations.
One of the major reasons for this is the absence of suitable simulation platform.

Networking simulation in LEO mega-constellations differs from well-known ones in terrestrial networks due to their sizeable spatiotemporal scale, high mobility, and orbital movements. 
These distinctive properties may pose significant challenges to traditional network simulators.

Until now, most simulators in community are combined with space simulators \cite{stk,gmat,savi} and network simulators \cite{ns3,mininet}, hence to obtain better functionality of visualization, celestial motion and network simulation. 
SNS3\cite{sns3} is an discrete simulator that combines NS3 with satellite communications link model, but it primarily supports simple functionality among GEOs without networking.
Handley \cite{handley2018delay} developed the platform through Unity 3D, which provides impressive visualization effects that intuitively show the path selection and delay of the end-to-end routing process, due to its emphasis on space simulation, it lacks the corresponding formation of the network simulation, and cannot satisfy the packet-level network simulation.
Benjamin\cite{kempton2021network} presents a network simulator for SNs with robust visualization features, leveraging OpenGL for enhanced graphics support. 
Celestial\cite{pfandzelter2022celestial} first introduced the virtualization technology into SN simulation/emulation, i.e., attaching each satellite with a micro virtual machine to emulate the real networking environment.
Hypatia\cite{kassing2020exploring} is developed based on Cesium\cite{cesium} in space simulation and based on NS3 for networking simulation, thus supporting packet-level simulation and satisfying visualization. 
Although it is not flexible in building scenarios, has limited visualization capabilities, and is no longer receiving updates, it is a valuable practice to implement complex visualization features through Web.
Overall, most simulation tools must possess several important qualities:
 1) effective visualization of space motion and communication processes; 
 2) flexibility to configure complex networking scenarios;
 and 3) scalability to support future multilateral requirements.




In this paper,
we introduce SNK, a performance simulation platform designed to assist constellation manufacturers and network operators in estimating and comprehending the attainable performance across various constellation options.

To demonstrate the effectiveness of SNK on routing algorithm and SN optimizing, 
we leverage SNK to evaluate and compare the performance of Dijkstra routing algorithm between cities and obtain insights on optimizing the constellation design to improve edge-to-edge network performance. 
The results of our evaluation indicate that the careful selection of a routing algorithm and network structure can significantly impact the communication process in terms of latency, path stretch, throughput and network capacity.

Overall, this paper makes two key contributions:
\begin{itemize}
	\item Introduction of SNK, a simulation platform that facilitates the profiling and understanding the network performance of mega-constellations under a diversity of network policies and structure. (\S.\ref{sec:method}).
	\item Utilization of SNK to reveal the different results under diverse network policies and structure, while also highlighting insights on optimizing constellation designs to improve network performance. (\S.\ref{sec:exp})
\end{itemize}

We are releasing our implementation of SNK available as open source to help future researchers validate their own applications and platforms. 
Our hope is that this will make the field of SN more accessible and provide a starting point for systems research in the community.

\section{The SNK platform}
\label{sec:method}

\subsection{System overview and workflow}

SNK is jointly developed in multiple languages and consists of four key modules: SNK-Scenario, SNK-Visualizer, SNK-Server and SNK-Analyzer.

The SNK-Scenario module serves the purpose of generating scenario data for \sn, including satellites, links, ground stations, mobile stations, and other relevant elements.
Developed as a web application rooted in Cesium, SNK-Visualizer goes beyond simple visualization of the constructed scenes. It dynamically showcases the networking processes, supports real-time adjustments of perspectives and layers, and facilitates interaction with the SNK-Server. The SNK-Server, a Python-based network simulator, establishes synchronous data transmission with SNK-Visualizer through its API subsystem. 
SNK-Analyzer functions as an analytical tool capable of visually representing statistical outcomes such end-to-end latency and path stretch rates.

\begin{figure}[t!]
	\begin{center}
		\includegraphics[width=1\linewidth]{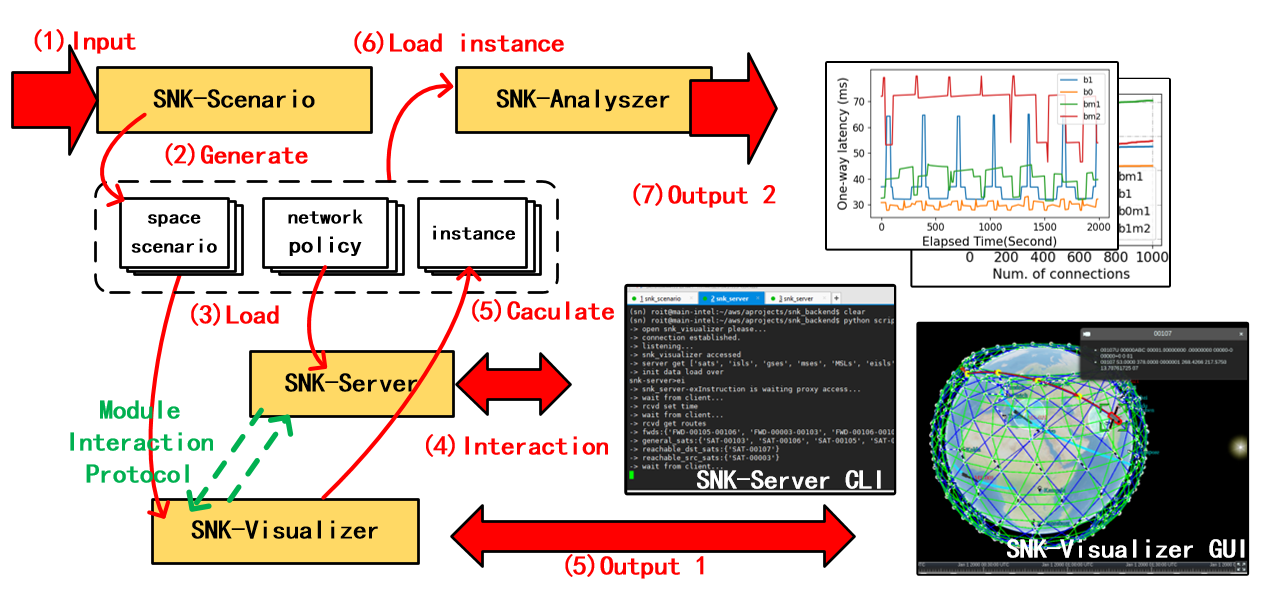}
	\end{center}
	\caption{The overview and workflow of SNK.} 
		\label{fig:teaser}
 \end{figure}

 \fig\ref{fig:teaser} shows the workflow of SNK. 
 The SNK takes a configuration file as input to generate scenario files, which describe the composition of constellation and networking.
 During runtime, the user interacts with SNK via the command line to load the created scenario file and initiate the simulation computation, and the simulation process can be observed in real-time by enabling the "Watch" mode. 
 At the end of the simulation, the generated instance file is loaded, 
 and performance metrics like network latencies and throughput are measured. These metrics are then used to compute and quantify the network performance.

\begin{figure*}[t!]
	\begin{center}
		\includegraphics[width=1\linewidth]{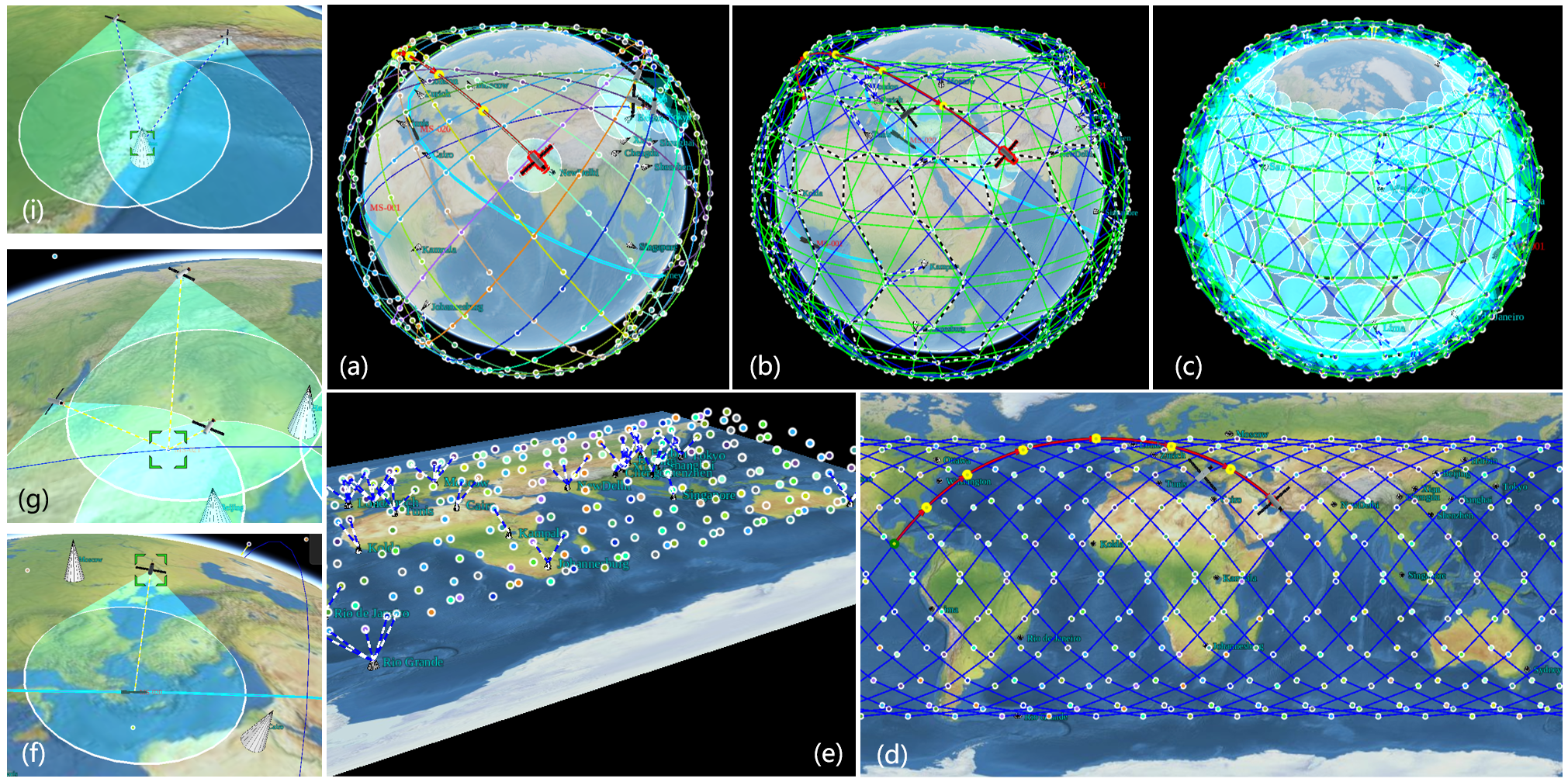}
	\end{center}
	\caption{The entities constructed by SNK-Scenario are displayed in SNK-Visualizer.} 
		\label{fig:sce}
 \end{figure*}

\subsection{SNK-Scenario}

SNK-Scenario consists of a suite of Python-based script programs designed to generate essential scenario entities required for simulation. 
These entities encompass a range of components including satellites (SATs), ground stations (GSes), and mobile stations (MSes), which may include operational satellites or airborne objects. It also encompasses links such as Inter-satellite links (ISLs), satellite-to-ground links (GSL), and mobile-station-satellite links (MSLs). Each of these entities is created via the execution of individual scripts.

SNK-Scenario offers remarkable flexibility, achieved through the utilization of distinct config.yaml files. This approach streamlines the creation of complex multi-layer \sns, such as Walker configurations with varying orbit arguments. 
A template for the config.yaml format is available in the supplementary documentation. 
Additionally, we provide a simplified script that simplifies scenario generation with a single bash command:

\begin{lstlisting}
	bash run.sh config.yaml
\end{lstlisting}
The construction of certain entities requires reference to other entities. Therefore, the recommended order of construction is as follows:

\begin{lstlisting}
SAT.py --> ISL.py--> eISL.py
MSes.py --> MSL.py
GSes.py --> GSL.py
\end{lstlisting}
Once the scenario is constructed, it is saved in the form of a \textit{*.sce} folder, with the general format being:
\begin{lstlisting}
scenario_name.sce/
  |- config.yaml
  |- {layer_name}_sats.czml
  |- gses.czml
  |- mses.czml
  |- {layer_name}_isls.czml
  |- {layer_name}_eisl.czml
  |- {layer_name}_gsls.czml
  |- {layer_name}_msls.czml
  |- {layer_name}_tle.txt
  |- {layer_name}_eisl.json
  |- {layer_name}_isls.json
  |- {layer_name}_gsls.json
  |- {layer_name}_msls.json
\end{lstlisting}

Among these, the *.czml format bears resemblance to a JSON file, primarily intended for interpretation by the web interface to create entities and is widely used in Cesium\cite{cesium}. Conversely, the *.json format functions as an abstract file, containing identifiers without essential details. Given the support for multi-layer networks, entities are tagged with numerical identifiers corresponding to network layers. In addition to furnishing satellite czml files, TLE data is also available to accommodate various  requirements.

\subsection{SNK-Visualizer}

\snkv is a web application developed using JavaScript. It runs in web browser, provides visual rendering capabilities for entities generated by \snks and enables seamless interaction with \snkser through the custmized protocol that base on WebSockets\cite{WebSocket}.

The constructed scenario and entities are depicted in  \fig\ref{fig:sce}, showcasing the outcomes of the \textit{config.yaml} script we provide.
\fig\ref{fig:sce} (a) shows a satellite constellation, a single-layer walker $\delta$ constellation with an inclination of $53^\circ$, consisting of 20 orbits, each containing 20 satellites.
\snkv allows the creation of various Inter-satellite Links (ISLs) (See \fig\ref{fig:sce} (b)), encompassing intra-orbit ISLs (blue), inter-orbit ISLs (green), and encountering ISLs (black and white striped).

Furthermore, \snkv supports showing routing paths between satellites or station nodes such as ground stations or mobile stations.
In (\fig\ref{fig:sce} (b)), the red lines represent routing paths between two satellites, established using the shortest-path routing algorithm, which takes the distance between satellites as edge weight. 
Satellites can be equipped with antennas with specific parameters, visualized by their coverage (See \fig\ref{fig:sce} (c)). 
This visualization can be employed in other \snkv modules for coverage analysis.
Ground stations within a satellite's coverage angle can establish GSL (blue and white striped lines in (\fig\ref{fig:sce} (i))). 
The 3rd-party satellites (\fig\ref{fig:sce} (g)) or aircrafts (\fig\ref{fig:sce} (f)) establish the MSL (yellow and white striped line) between the satellite and the mobile station.
\snkv offers two perspectives - 3D perspective (\fig\ref{fig:sce} (a,b,c)) and 2D perspective (\fig\ref{fig:sce} (e,d)) - facilitating a comprehensive analysis of scene entities.

\subsection{SNK-Server}

\snkser is developed by Python and is primarily designed with network simulation and interaction with SNK-Visualizer. 
It represents the most intricate module within the whole system, encompassing three major subsystems: the CLI subsystem, the API subsystem, and the procedure subsystem.
The CLI subsystem functions as the command-line interface, predominantly responsible for user interaction and operational logic. Our current emphasis is on providing more insights into the API subsystem and procedure subsystem.

\subsubsection{API subsystem}

\snkser interacts with modules or external modules through API subsystem using the Module Interaction Protocol (MIP), a command and data transmission protocol developed based on WebSocket, facilitating the data sharing and simulation consistency.

\subsubsection{Procedure subsystem}

The Procedure subsystem plays a pivotal role in orchestrating the scheduling of each procedure within the \snkser. The execution of procedures serves two distinct purposes:
\begin{itemize}
	\item Data acquisition for quantitative analysis. 
	One primary objective of procedure execution is to acquire data periodically. 
	Acquired data is a key factor used to evaluate the performance of a communication policy or network, and stands as the foremost mission of the Server.
	\item Observation for qualitative analysis. By visualizing the communication process in a dynamic network, it is easy for users to debug and improve the networking policy.
\end{itemize}
Within the such subsystem, we offer several procedures as following:

\textbf{The edge2edge procedure.}
It performs the communication process between edge nodes (GSes and MSes) defined in network policy.
Upon its activation, the networking policy and a timestamps are initialized and it  runs as the following steps:
\begin{itemize}
	\item[(1)] When \snkser starts up, it first gets the simulation timestamp from network policy file and sends the MIP command "\textit{set time}" to \snkv for consistency.
	\item[(2)] 
	\snkser sends request commands to get the space scenario information in \snkv, including network topology status, satellite positions, and other pertinent information.
	\item[(3)] 
	Based on the acquired data, \snkser starts to calculate the routing path between the source and destination satellite sets. 
	The source and destination satellite sets are the satellites covering the source edge node and destination edge node respectively.

	\item[(4)]  Upon reaching the predefined end timestamp, the procedure is halted.
	\item[(5)] The procedure information is then preserved in \textit{*.ins} files as the instance data.
\end{itemize}

\textbf{The conTest procedure.}
It performs the communication process between any two of satellites nodes, serving the purpose of evaluating the network policy or space scenario.
Upon its activation, the routing algorithm and a time list are initialized and it  runs as the following steps:
\begin{itemize}

	\item[(1)] When \snkser starts up, it first gets the simulation timestamp from network policy file and sends the MIP command "\textit{set time}" to \snkv for consistency.
	\item[(2)] 
	\snkser sends request commands to get the space scenario information in \snkv, including network topology status, satellite positions, and other pertinent information.
	\item[(3)] 
	Based on the acquired data, \snkser starts to calculate the routing path between any two satellites in the scenario under built-in algorithms.
	\item[(4)] if the number of paths exceeds a predefined threshold $N$, the procedure is interrupted, and the timestamp is updated.
	\item[(5)]  Upon reaching the predefined end timestamp, the procedure is halted.
	\item[(6)] The procedure information is then preserved in \textit{*.ins} files as the instance data.
\end{itemize}


In addition to the first two procedures, we also provide: i) \textbf{asyncReplay procedure}, a procedure that can convert the event list and packet list generated by external simulation programs into animation data and visualize it asynchronously; ii) \textbf{replay procedure}, a procedure based on the obtained \textit{*.ins} file to replay the process of it and iii) \textbf{exInstruction} a procedure that supports joint debugging with external programs by API subsystem.







\section{Simulation cases}
\label{sec:exp}

In this section we demonstrate the effectiveness of SNK on characterizing and understanding emerging LEO mega-constellation systems.

\subsection{The Local View}
This Simulation involves routing analysis between fixed edge stations (representing ground stations or user terminals) and mobile edge stations (representing functional satellites or aircraft).
The scenario is constructed as shown in \tab\ref{tab:sce1}, and the stations are listed in \tab\ref{tab:stations}.
\begin{table}[htbp]
	\caption{Scenario information.}
	\label{tab:sce1}
	\centering
	\scalebox{0.96}{
		\begin{tabular}{c|ccc}
			\toprule[2pt]
			\vspace{2pt}
Symbols & description    & value     \\ \hline
$T/P/F/i$     		&parameters of constellation  &$400/20/0/53^\circ$   \\
$h$      	& altitude of orbits   & 1000km  \\
 - & eISL building threshold\cite{eISL} & 2000km\\
-   		& ISL mode   & *Grid  \\
$\alpha$	& steering angle of beams   & $45^\circ$   \\ 

-		& Simulation time       & 2000s \\
-		& simulation time delta       & 10s \\

\toprule[2pt]
\end{tabular}
}

\end{table}
\begin{table}[htbp]
	\caption{Edge stations.}
	\label{tab:stations}
	\centering
	\scalebox{0.96}{
		\begin{tabular}{c|ccc}
			\toprule[2pt]
			\vspace{2pt}
Edge Station Name & ID     & Location $\backslash$ Trajectory     \\ \hline
Harbin     		& GS-0000   & 45N,127E   \\
London      	& GS-0001   & 51N,0E  \\
Chengdu   		& GS-0002   & 31N,102E  \\
San Francisco	& GS-0003   & 38N,122W   \\
Shanghai		& GS-0004  	& 30N,122E        \\
Johannesburg	& GS-0005  	& 26S,28E     \\ \hline
MS-000		& MS-000 	    & 0N,24W  $\rightarrow$   0N,20E    \\
MS-020		& MS-020	  	& 50N,0E    $\rightarrow$  50S,180E\\ 
MS-021		& MS-021	  	&   - \\ 

\toprule[2pt]
\end{tabular}
}

\end{table}

\subsubsection{Fixed edge station routing}

\begin{figure}[t!]

	\subfigure[Path latency over time between cities under different routing strategies.]
	{
	\begin{minipage}{0.98\linewidth}
	\includegraphics[scale=0.55]{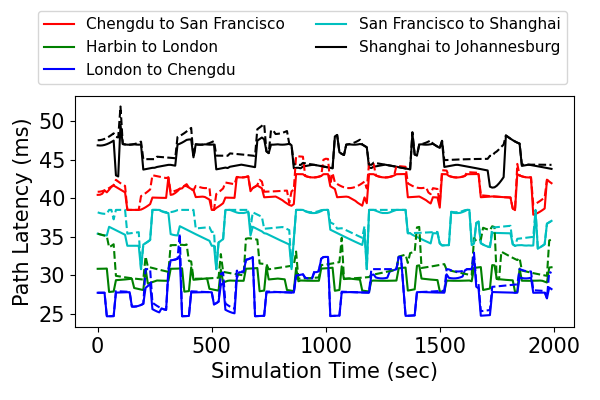}
	\end{minipage}
	}
	
	\subfigure[The path stretch and evolution between city pairs.]
	{
	\begin{minipage}{.98\linewidth}
	\includegraphics[scale=0.43]{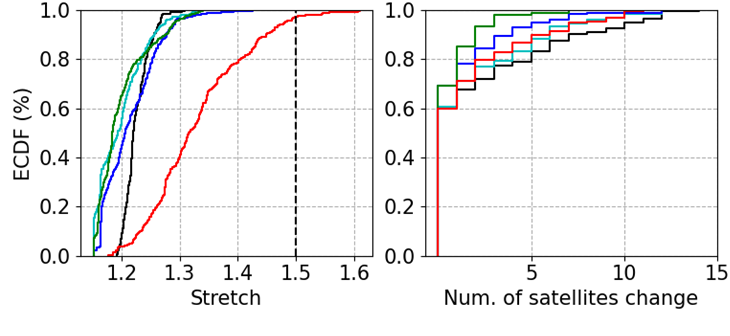}
	\end{minipage}
	}
	   \caption{Space network structures.} 
			   \label{fig:cities}
   \end{figure}

 \begin{figure}[t!]
    \begin{center}
        \includegraphics[width=.7\linewidth]{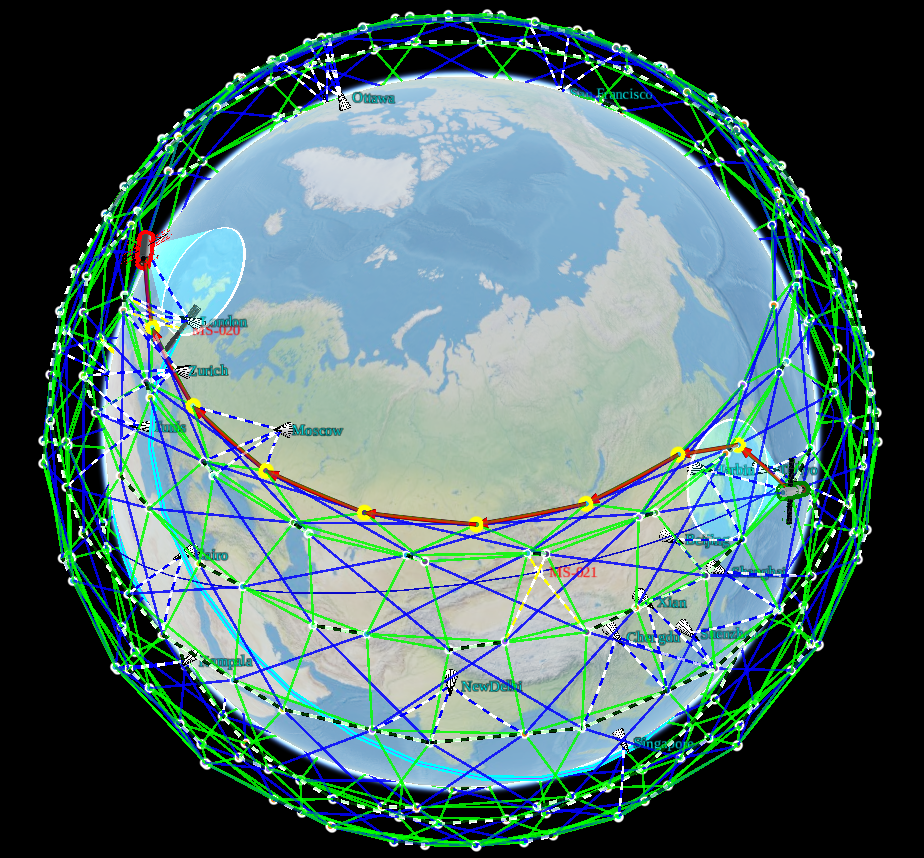}
    \end{center}
    \caption{The routing path can be visualized in \snkv in real time when the enable "Watch" mode.} 
        \label{fig:har2lon}
 \end{figure}

SN can effectively reduce network latency between cities and offers an alternative solution to submarine cables, therefore we will analyze the routing between fixed edge stations located in large cities with the \sn.
When the scenario is constructed via \snks and the \textit{edge2edge} process is executed, \snkser generates the \textit{*.ins} directory and retrieves routing information from \snka.

\fig\ref{fig:cities} (a) displays the path latency under two different routing strategies: least-hop (dashed lines) and shortest-path (solid lines) routing, for various city pairs over a 2000-seconds period.
We have provided basic shortest path algorithms in SNK, and more complex routing algorithms can be constructed with simple development.
In \fig\ref{fig:cities} (b) on the left, the stretch distribution of the routing path is depicted. This metric measures the ratio of the routing path distance over the geographic distance, indicating the path detour. 
A smaller stretch indicates a path closer to the geographic arch, indicating a more efficient path.
If $stretch \leq 1.5$, it indicates that the routing path in \sn is better than the optical fiber direct connection between two cities\cite{kassing2020exploring}.
It is clear that the space network outperforms the terrestrial network in most routing strategies.
On the right side of \fig\ref{fig:cities} (b), the number of satellite changes in the path over time is shown, with smaller values indicating stable routing.
In addition, in order to facilitate debugging during the development of routing algorithms, SNK has the capability to display the routing paths in real time in "Watch-enabled" mode, as is shown in \fig\ref{fig:har2lon}.

\subsubsection{Mobile station routing}
\begin{figure}[t!]

	\subfigure[Mobile stations include aircraft (MS-001, MS-020) or functional satellite (MS-021).]
	{
	\begin{minipage}{0.9\linewidth}
	\includegraphics[scale=0.25]{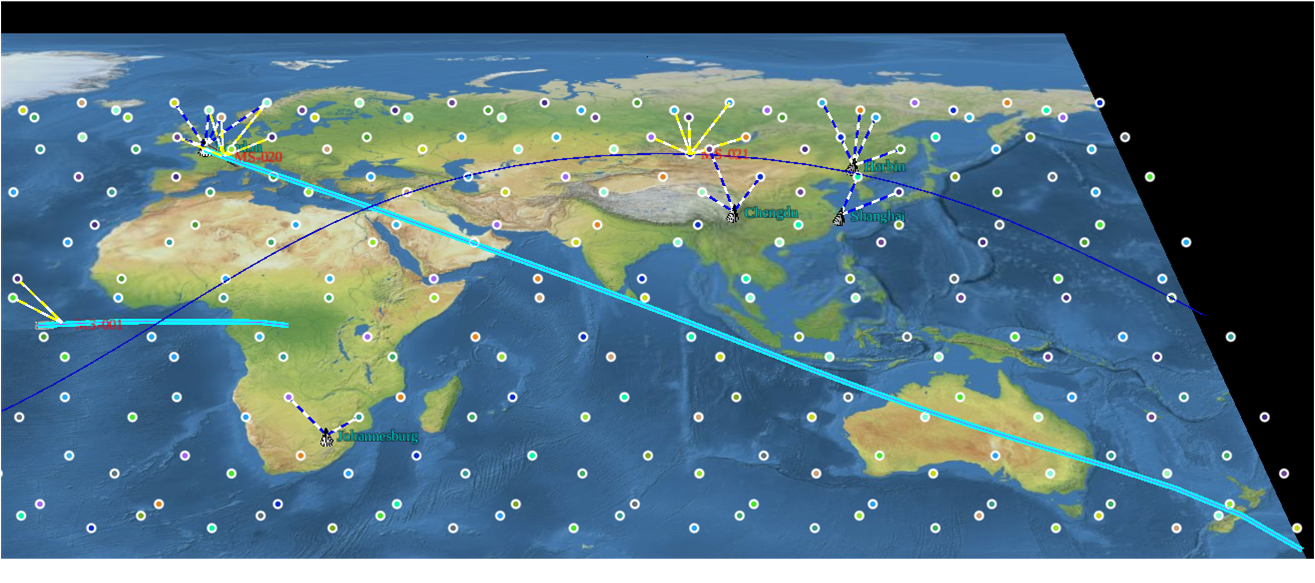}
	\end{minipage}
	}
	\subfigure[Path latency of mobile edge stations.]
	{
	\begin{minipage}{1\linewidth}
	\includegraphics[scale=0.5]{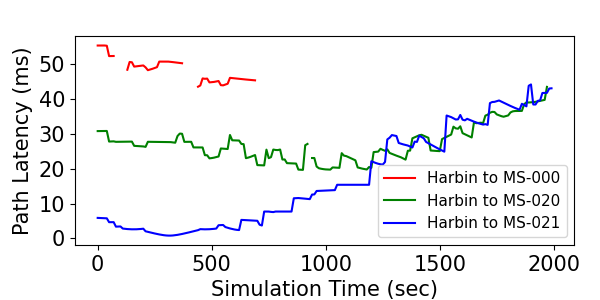}
	\vspace{5pt}
	\end{minipage}
	}
	   \caption{Space network structures.} 
			   \label{fig:ms-lat}
   \end{figure}
SNK also supports to simulate the routing procedure between mobile edge nodes with a customized trajectory, such as aircraft or 3rd-party satellites such as Earth observation satellites.
In this case, we constructed multiple mobile edge stations with different trajectories and evaluated the path latency between them and Harbin under different routing algorithms, as shown in \fig\ref{fig:ms-lat} (a).
The thinner blue path is the orbit of 3rd-party satellites and thicker ones are the paths of the aircraft.
The dashed yellow lines stands for mobile-station satellite links (MSLs).
\fig\ref{fig:ms-lat} (b) shows the path latency of MSes. Note that there are gaps in the lines of MS-000 and MS-020 because there is not always an accessible satellite during the movement of these mobile stations.

 \subsection{The Global View}
 
 Existing constellations are primarily designed for coverage optimization or conflict avoidance,
 however, it is evident that designing systems with improved network performance in SNs is of paramount importance as it directly impacts the availability and feasibility of such systems.
 Therefore, SNK provides functions to quantify the network performance under different network structures and density, including global latency, stretch, network capacity and throughput.
 This capability supports the design of SNs and included in this simulation case. The scenario configuration is listed in \tab\ref{tab:sce2}.

 \begin{figure}[t!]
    \begin{center}
        \includegraphics[width=1\linewidth]{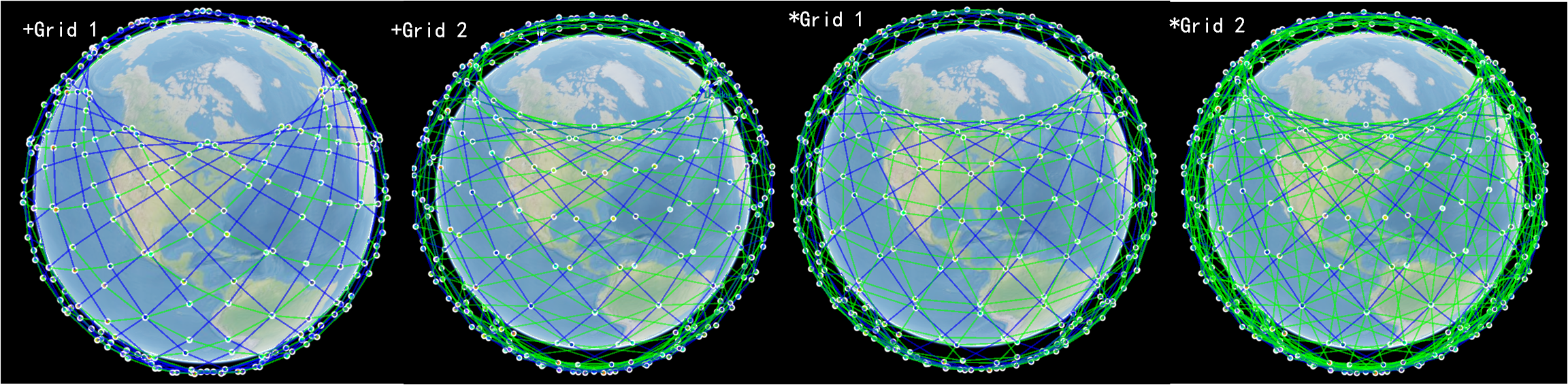}
    \end{center}
    \caption{Different configurations build vastly different network structures} 
        \label{fig:structures}
 \end{figure}
 \begin{table}[htbp]
	\caption{Scenario configuration of global analysis.}
	\label{tab:sce2}
	\centering
	\scalebox{0.96}{
		\begin{tabular}{ccc}
			\toprule[2pt]
			\vspace{2pt}
Description    & value     \\ \hline
Constellation scales   &\{$10^2,20^2,30^2,40^2$\}   \\
	 Constellation structures   & \{+Grid 1, +Grid 2, *Grid 1, *Grid 2   \}   \\ 
Max num of connections & 1000 \\
Simulation time       & 1000s \\
 Simulation time delta       & 100s \\

\toprule[2pt]
\end{tabular}
}

\end{table}

\subsubsection{Structure analysis}

\begin{figure}[t!]

	\subfigure[Throughput.]
	{
	\begin{minipage}{0.6\linewidth}
	\includegraphics[scale=0.48]{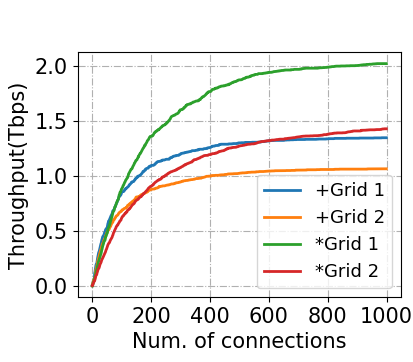}
	\end{minipage}
	}\subfigure[Latency.]
	{
	\begin{minipage}{.5\linewidth}
	\includegraphics[scale=0.5]{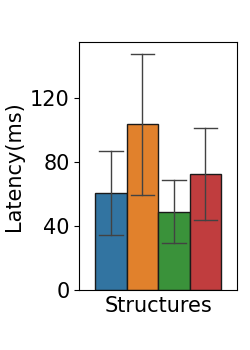}
	\end{minipage}
	}
	   \caption{Throughput and latency distribution in different network structures.} 
			   \label{fig:net}
   \end{figure}

 The network structure analysis aims to evaluate the network performance of a constellation under a specific network structure, providing valuable insights for proposing robust network structures.
  \fig\ref{fig:structures} shows the different structures under configuration files in SNK which are denoted as +Grid 1, +Grid 2, *Grid 1 and *Grid 2 respectively and represent 4 and 6 adjacent ISLs per satellites\cite{bhattacherjee2019network,eISL}.
  In the above structure, the capacity of each link can be easily obtained based on the free space loss, and then the network throughput is calculated by the maximum flow algorithm.
  \fig\ref{fig:net} illustrates the throughput and latency distribution in different network configurations, providing insights into the impact of network structure on these performance metrics.

\subsubsection{Density Analysis}

\begin{figure}[t!]

    \begin{center}

        \subfigure[Network throughputs.]
        {
            \begin{minipage}{0.52\linewidth}
                \includegraphics[scale=0.47]{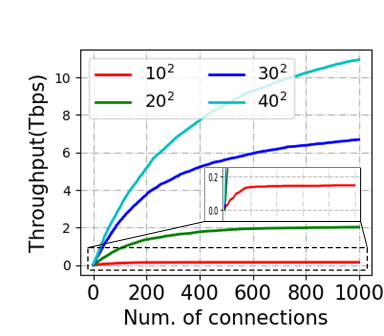}
            \end{minipage}
        }\subfigure[Network capacity.]
        {
            \begin{minipage}{.4\linewidth}
                \includegraphics[scale=0.47]{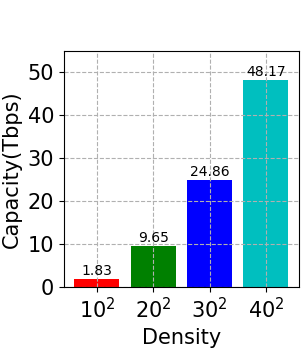}
            \end{minipage}
        }
    \end{center}

    \caption{Network capacity and throughput under shortest-path routing in structures with density $10^2,~20^2,~30^2,~40^2$.}
    \label{fig:thp_scale}
\end{figure}

Constellation networks of different density perform differently, with higher density networks exhibiting higher throughput and lower latency.
\fig\ref{fig:thp_scale} shows the throughput and capacity of networks with different density under *Grid 1 structure. 
It demonstrates that as the network size increases, the capacity and throughput are increasing significantly.
However, it's important to note that since the routing algorithm relies on basic shortest-path without a load balancing mechanism, the capacity utilization remains around 25\%.

\section{Limitations and Future works}

\textbf{Optimizing concurrent processing to enhance smoothness.}
As the scale of space scenario and the complexity of network grow, the simulation process becomes less fluid.
  In the future, SNK will be further updated to achieve smoother simulations by optimizing the concurrency process.

\textbf{Introducing virtualization technology to SNK for high-fidelity emulating.}
Unlike other recent works that focus on the network emulating\cite{mininet} and space network emulating\cite{pfandzelter2022celestial}, 
SNK relies on model-based estimation for networking performance results. To improve fidelity, SNK's latest update combines virtualization techniques to emulate realistic networking environments by isolating node entities in network namespaces and connecting nodes using a novel link emulator\cite{netem,petersen2021dynamic}. This enhancement aims to enable system-level emulation and facilitate more accurate performance comparisons with real systems like Starlink or OneWeb.

  \section{Conclusion}

  This paper presents SNK, a novel simulation platform that enables constellation manufacturers and network operators to estimate the achievable network performance under a variety of constellation options. 
  SNK allows user to build large-scale complex scenarios with configuration files by typing a single bash command, and evaluate or visualize the communication process.
Leveraging SNK, we evaluate and compare the performance of shortest-path and least-hop routing algorithm and obtain the insights on network optimization for mega-constellations;
This platform will be open-sourced to help future researchers validate their own applications and platforms.



\bibliographystyle{IEEEtran}
\bibliography{ref}
\end{document}